\documentclass[aps,prl,twocolumn,groupedaddress, superscriptaddress,longbibliography]{revtex4-1}
\usepackage{graphicx}
\usepackage{amsmath}
\usepackage{url}
\usepackage[normalem]{ulem}
\usepackage{enumitem}

\usepackage[colorlinks,bookmarks=false,citecolor=blue,linkcolor=red,urlcolor=blue]{hyperref}

\begin{document}

\title{Quantum memory at an eigenstate phase transition in a weakly chaotic model}
\author{M. R. Lambert}
\affiliation{Theoretical Division, Los Alamos National Laboratory, Los Alamos, New Mexico 87545, USA }
\affiliation{Department of Physics and Astronomy, University of California Riverside, Riverside, California 92521, USA}
\author{Shan-Wen Tsai}
\affiliation{Department of Physics and Astronomy, University of California Riverside, Riverside, California 92521, USA}
\author{Shane P. Kelly}
\email[Corresponding author:~]{shakelly@uni-mainz.de}
\affiliation{Institut f\"ur Physik, Johannes Gutenberg Universit\"at Mainz, D-55099 Mainz, Germany}
\affiliation{Kavli Institute for Theoretical Physics, University of California, Santa Barbara, CA 93106-4030, USA}
\date{\today}

\begin{abstract}
    We study a fully connected quantum spin model resonantly coupled to a small environment of non-interacting spins, and investigate how initial state properties are remembered at long times.
    We find memory of initial state properties, in addition to the total energy,  that are not conserved by the dynamics.
    This memory occurs in the middle of the spectrum where an eigenstate quantum phase transition (ESQPT) occurs as a function of energy.
    The memory effect at that energy in the spectrum is robust to system-environment coupling until the coupling changes the energy of the ESQPT.
    This work demonstrates the effect of ESQPT memory as independent of integrability and suggests a wider generality of this mechanism for preventing thermalization at ESQPTs.
\end{abstract}

\maketitle
\newlength{\figsize} 
\setlength{\figsize}{1.0\columnwidth}

Investigations into the past are often inhibited by the natural tendency of the world to forget and are sometimes 
%must accept that 
faced by the possibility that
%or "by the fact that"?...
certain aspects of past events are unknowable.
Other times, scientists benefit from this loss of memory because it allows them to neglect the details of how an object of study was formed.
Thermal equilibrium is an examples of this and allows for models that require only a few macroscopic properties such as temperature and pressure. 
Thus, the quest for a general set of principles to understand when systems forget has been actively pursued throughout the 20th century~\cite{brin_stuck_2002,RahulReview,Abanin2019}.
In the past 20 years, the question of memory has become increasingly important and relevant for the dynamics of quantum systems due to new experiments that are effectively decoupled from a thermal bath~\cite{Bloch2008,Jorg2015,Abanin2019,zhang2017,muniz2020,norcia2018b,davis2019}, and the technological push for quantum computing which relies on maintaining memory of quantum information~\cite{knill_quantum_2005,Roffe2019,ladd_quantum_2010}.

In quantum systems decoupled from the environment, the Eigenstate Thermalization Hypothesis (ETH) has been the guiding principle to understand thermalization and loss of memory.
Such a hypothesis has been verified directly in a number of models  by both numerical experiments and indirectly in physical experiments~\cite{Shrednicki1994,Deutsch1991,PhysRevLett.108.110601,Deutsch_2018,Alessio2016}.
Furthermore, a classification of systems that do not thermalize has also been developing.
In quantum integrable models~\cite{Clark2011,Rigol2007,kinoshita_quantum_2006}, the existence of a set of local conserved charges invalidates the above assumptions and yields, at late times, memory of those conserved charges.
In many-body localization~\cite{RahulReview,Abanin2019}, a similar phenomenon occurs in which an emergent set of local conserved charges appear.

In addition, some systems have been found to generically thermalize, but host a set of scarred states~\cite{serbyn_quantum_2021,buca_non-stationary_2019} that do not follow ETH.
Such systems often show dynamics with persistent oscillations and long-time states not described by thermal distributions.
Recently, some of the authors of this article found a similar phenomena occurring at a thermal phase transition~\cite{Kelly2020}.
In that work, quantum memory of non-conserved initial state properties persists at late times for initial states quenched at the energy of the thermal phase transition, while memory of the same initial state property is lost away from the transition.

In this work, we investigate the generality of this memory effect occurring at a thermal phase transition.
The generality was proposed in the original work~\cite{Kelly2020} by a simple argument in the context of ETH.
Specifically, in both that work and in ETH, the question of late time memory considers the dynamics of a Hamiltonian $H$, evolving an initial state $\left|\psi(\left\{\theta_i\right\})\right>$ with parameters $\theta_i$, and asks how the late time average dynamics of a local observable  $O$:

\begin{eqnarray} \label{eq:latetime}
    \overline{O}(\{\theta_i\})=\lim_{T\rightarrow \infty} \frac{1}{T} \int_0^T dt O(t)=\sum_n \left|c_n\right|^2 \left<n\right|O\left|n\right>
\end{eqnarray}
depend on $\left\{\theta_i\right\}$, where in Eq.~\ref{eq:latetime}, $\left|n\right>$ is an eigenstate of $H$, $c_n=\left<n|\psi(\theta_i)\right>$, and the sum is over all eigenstates $n$.
ETH explains the observed fact that, for most quantum quenches, memory is lost by making a hypothesis for $c_n$ and $\left<n\right|O\left|n\right>$:
\begin{enumerate}
    \item Initial states have sufficiently small energy variance such that $c_n$ is narrowly distributed around a single energy.
    \item Eigenstate expectation value of local observables such as $O$ depend only on the energy eigenvalue, $e_n$, and $\left<n\right|O\left|n\right>=O(e_n)$ is a smooth function.
\end{enumerate}
Under these assumptions, a typical eigenstate $\left|m\right>$ with $e_m\approx \left<\psi(\theta_i)\right|H\left|\psi(\theta_i)\right>$ will accurately reproduce the long time average $\overline{O}(\{\theta_i\})\approx \left<m\right|O\left|m\right>$.
Thus, when these assumptions of ETH are true, the initial state information encoded in $c_n$ is lost to the late time observable and only the initial state energy is remembered at long times. 
Finally, if the system is expected to thermalize, and $\bar{O}(\{\theta_i\})$ is expected to match a thermal ensemble, then one expects eigenstate observables $\left<m\right|O\left|m\right>$ to match observables for thermal ensembles with temperature chosen to match the energy of the initial state. Such an expectation, deduced from the assumptions above, is the ETH.

If a thermal phase transition is expected to occur and ETH is true, then symmetry breaking and the singularities due to the transition must show up in $O(e_n)$.
This is because, as the temperature (or energy) is reduced at a thermodynamic phase transitions, the late time observables, $\overline{O}(\{\theta_i\})\approx \left<m\right|O\left|m\right>$, will show phase transition singularities~\cite{kardar_2007} as a function of energy.
But if this is true, then assumption 2) of ETH is not valid, and the singularity can make the late time observable sensitive to initial conditions.
Thus there appears to be a contradiction between thermalization at thermal phase transitions and ETH.
We imagine that the resolution of this contradiction could occur in one of three ways:
\begin{enumerate}[label=\Alph*)]
    \item The thermal phase transition is reflected directly in the eigenstate observables $O(e_n)$, in what is known as an excited-state quantum phase transition (ESQPT)~\cite{Cejnar_2021,CAPRIO20081106}, and singularities  in the eigenstate observables can cause late time memory and the breakdown of thermalization. In this case, assumption 2) for ETH is not valid,  and both thermalization and loss of memory do not occur.
    \item Loss of memory occurs because eigenstate observables do not contain singularities that can produce memory effects. Given assumption 1) above, this necessarily implies late time observables, $\overline{O}(\{\theta_i\})\approx \left<m\right|O\left|m\right>$, and the eigenstate observables $\left<m\right|O\left|m\right>$ do not match the singular observables of thermal ensembles. In this case the ETH is false, the late time steady state does not show a phase transition and while loss of memory occurs, thermalization does not.
    \item Singularities are present in $O(e_n)$, but it is difficult to prepare initial states that are sensitive to it.  If this is the case, the ETH is not satisfied and whatever mechanism preventing that sensitivity is a currently unknown mechanism for thermalization.
\end{enumerate}
The first option was found in a model with a Z$_2$ ESQPT that leads to late time memory of initial conditions in local observables~\cite{Kelly2020}. 
The model in that work is both classical and quantum integrable~\cite{Raghavan1999,MORITA2006337}, and thus to appreciate the relevancy of the above contradiction, one should use the generalized eigenstate thermalization hypothesis~\cite{Clark2011, vidmar_generalized_2016}.
     This hypothesis has statements similar to those of 1) and 2) of ETH, but with energy replaced by a sufficiently large set of local conserved charges of the integrable model~\cite{schmitt_observations_2021}.
     This complication is irrelevant to the model in Ref.~\cite{Kelly2020} for which the classical phase space is two dimensional and integrability only requires conservation of energy.
     Thus, in Ref.~\cite{Kelly2020}, assumptions 1) and 2) were confirmed for dynamics away from the ESQPT, while again posing the contradiction discussed above for dynamics at the transition. 
 The primary difference between generalized ETH and ETH~(and consequentially the model in Ref.~\cite{Kelly2020} and a chaotic model), is that ETH also hypothesizes Wigner Dyson level statistics to reflect the fact that the dynamics of non-integrable models~\cite{Deutsch_2018} are chaotic.
     The model in Ref.~\cite{Kelly2020} does not satisfy these additional assumptions and while the late time memory observed can not be explained by the conserved charges, the existence of late time memory is from one perspective unsurprising due to the lack of chaos. 
     This motivates us to study the generality of the above arguments to a model displaying chaotic dynamics at the ESQPT such that any form of late time memory is traditionally expected to be lost. 

%While the model in that work was both classical and quantum integrable~\cite{Raghavan1999,MORITA2006337}, the argument for memory due to the phase transition singularities appears to be more general.

Such generality might also be anticipated by similar results studying thermalization near phase transitions.
In Ref.~\cite{Brenes2020} the quantum Fisher information is shown to be particularly sensitive to the non-thermal nature of a closed quantum systems at a thermal phase transition.
While in the Kibble-Zurek mechanism~\cite{delcampo2014}, critical slowing down near a phase transition leads to the freeze out of domain walls, and the number of domain walls remembers, in effect, the rate at which the system crossed the phase transition.
Furthermore, recent work has found that long-range interacting models, which often host ESQPTs~\cite{Cejnar_2021}, generically fail strong ETH~\cite{sugimoto_eigenstate_2021}.
Finally, recent work studying chaos quantifiers, such as the out of time ordered correlators, observe unexpected behaviour near eigenstate phase transitions~\cite{bhattacharjee2022,xu_does_2020,varikuti_out--time_2022}.

In these examples, late time quantum memory is not investigated and Ref.~\cite{Brenes2020} studies a non-local observable where ETH is not expected to apply.
Furthermore, the additional assumptions of ETH on the chaotic level statistics did not apply in the integrable model of Ref.~\cite{Kelly2020} which found case A) above. 
Thus the full implications of the apparent contradiction discussed above is not yet understood.
Therefore, in this work, we study the model in Ref.~\cite{Kelly2020} coupled to an extensive number of non-interacting spins, which act as an environment, and ask the same question regarding quantum memory at the excited state phase transition.

First we discuss the model, and show that the coupling to the non-interacting environment is an integrability-breaking parameter that, even at weak coupling, creates chaotic dynamics near the ESQPT.
We then find that for weak and moderate coupling strength, the late time quantum memory persists and even induces a type of proximity effect where the environment degrees of freedom are also sensitive to the initial state parameters of the system.
We also find that at large couplings, the memory effect is lost and argue that this is because at large couplings, the ESQPT is shifted to lower energies than for the initial states we considered.
These results are found via a combination of exact diagonalization~\cite{10.21468/SciPostPhys.7.2.020,SciPostPhys.2.1.003}, classical chaos analysis~\cite{DynamicalSystems.jl-2018} and the Truncated Wigner Approximation (TWA)~\cite{POLKOVNIKOV20101790}.

\section{Model and Classical Dynamics}

We consider a fully connected spin model fully coupled to a non-interacting environment of spins evolving according the Hamiltonian
\begin{eqnarray}
    H= H_s+H_e+H_v,
\end{eqnarray}
where
\begin{eqnarray}
    H_s&=& -\frac{h_s}{2}\sum_{i=1}^{N} \sigma^x_{s,i}+\frac{\Lambda}{4N}\sum_{ij}\sigma^z_{s,i}\sigma^z_{s,j} \\ \nonumber
    H_e&=& -\frac{h_e}{2}\sum_{i=1}^{N} \sigma^x_{e,i}, \\ \nonumber
    H_v  &=& \frac{V}{8N}\sum_{ij} \sigma^z_{s,i}\sigma^z_{e,j},
\end{eqnarray}
$N$ is both the number of system spins and the number of environment spins, and $\sigma_{s,i}^{z(x,y)}$ and $\sigma_{e,i}^{z(x,y)}$ are the system and environment Pauli spin operators.
Throughout this paper, we consider the environment in resonance with the system such that $h_e=h_s=h$ and the system interaction strength is fixed to $\Lambda/h=10$.
Note, that the environment we consider has the same size as the system, and thus non-Markovian effects are expected to be relevant and a master equation approach invalid.

Both the environment and the system exhibit a permutation symmetry, allowing the dynamics to be described by collective spin variables $\vec{S}_s=\frac{1}{2}\sum_i\vec{\sigma}_{s,i}$ and $\vec{S}_e=\frac{1}{2}\sum_i\vec{\sigma}_{e,i}$ with Hamiltonian:
\begin{eqnarray}\label{eq:collectiveHam}
    H=-hS_s^x-hS_e^x+\frac{\Lambda}{N}(S^{z}_s)^2+\frac{V}{2N}S^{z}_sS^{z}_e.
\end{eqnarray}
Such a model has a Z$_2$ symmetry corresponding to changing the sign of both system and environment operators: $S^z_s\leftrightarrow -S^z_s$ and $S^z_e\leftrightarrow -S^z_e$.
While at $V=0$ it has a Z$_2 \times$Z$_2$ symmetry corresponding to changing the sign of either the system or environment operators: $S^z_s\leftrightarrow -S^z_s$ or $S^z_e\leftrightarrow -S^z_e$.
In the decoupled limit, the dynamics of the system is integrable~\cite{MORITA2006337} and shows an ESQPT~\cite{CEJNAR2009210,PhysRevA.80.032111,PhysRevA.85.044102,PhysRevA.94.012113,PhysRevB.104.085105} at which the system breaks the Z$_2$ symmetry $S^z_s \rightarrow -S^z_s$, and is capable of memory in the collective observables $\left<\vec{S}_s\right>$ at the ESQPT as discussed above. Note that while the collective observables have extensive scaling, they are directly related to averages of local observables $\left<\vec{S}_s\right>= \sum_i\left<\vec{\sigma}_{s,i}\right>/2$. Since ETH applies to the local observables $\vec{\sigma}_{s,i}$, it therefore also applies to extensive sums of local observables~\cite{rigol_thermalization_2008}, such as $\vec{S}_s$.

\begin{figure}[]
    \centering
    \hspace*{-0.5cm}
    \includegraphics[width=\figsize]{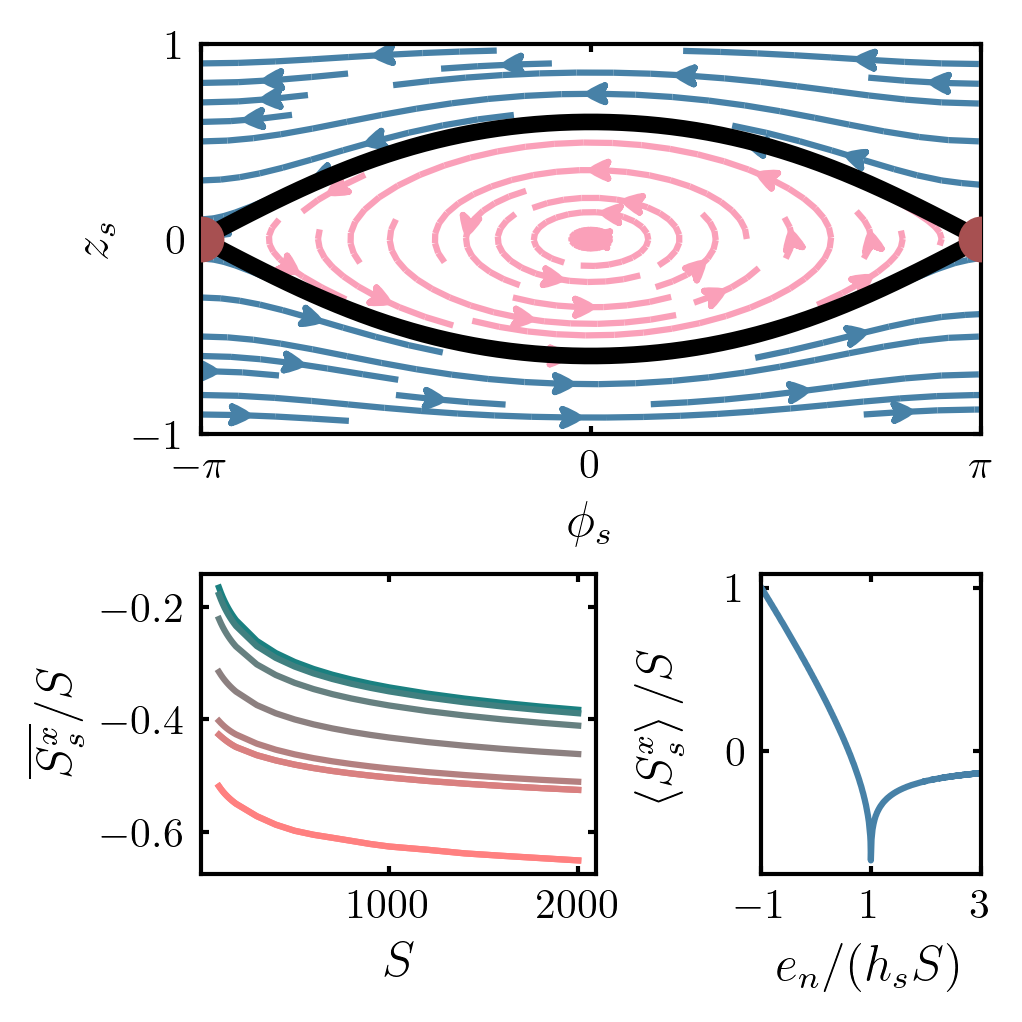}
    \caption{\textit{Top Panel}: This figure shows the classical trajectories of the collective spin variables $(\phi_s, z_s)$ for the semi-classical limit of the system Hamiltonian $H_s$. The trajectories with energy at the ESQPT are highlighted in black, and form two separatrices that meet at the unstable fixed point at $(\phi_s,z_s)=(\pi,0)=(-\pi,0)$. The separatrix separates trajectories with sign changing $z_s$ (marked in bright pink), and trajectories that do not change sign (marked in dark blue). \textit{Bottom Left Panel}: This figure shows the long time average of $x$-polarization of the collective spin, $S^x_s$ for $V=0$ as a function of the system spin size $S$, and it demonstrates the memory of the initial phase $\phi_s(t=0)$ at the ESQPT.  The initial states that produce these long time averages all have energy at the ESQPT $e_n/(h S)=1$, but with different initial phases $\phi_s(t=0)$ from $\phi_s(t=0)=0$ (darkest blue) to $\phi_s(t=0)=\pi$ (brightest pink). \textit{Bottom Right Panel}: This figure shows the expectation value of $S^x_s$ in the eigenstate $\left|n\right>$ at energy $e_n$. At the energy of the ESQPT, $e_n/(hS)=1$, $\left<S^x_s\right>$ approaches $-1$ non-analytically.
    }
    \label{fig:classicalOG}
\end{figure}

The ESQPT and the long-time memory at $V=0$ were discussed in Ref.~\cite{Kelly2020}, and are most easily understood close to the thermodynamic limit when $N>>1$.
In this limit, a semi-classical dynamics is valid and tracks the evolution of the collective spin mode $\vec{S}_s$ with size $S=N/2$, via the classical conjugate variables ($\phi_s, z_s$), where $z_s$ is the projection of the collective spin onto the $z$-axis $z_s$, and $\phi_s$ is the angle of the collective spin from the $+\hat{x}$-axis in the $x$-$y$ plane.
The evolution of these classical degrees of freedom can be solved exactly~\cite{Raghavan1999}, and the resulting trajectories are shown in Fig.~\ref{fig:classicalOG}.
The eigenstate phase transition that occurs at $\left<H_s\right>/(h S)=E(\phi_s, z_s)=1$~\cite{CEJNAR2009210} is reflected in the classical dynamics as a separatrix that emanates out from an unstable fixed point at $\langle\vec{S}_s\rangle= -S\hat{x}$, i.e., at $(\phi_s, z_s)=(\pm\pi,0)$.
The separatrix separates symmetric trajectories, with $E<1$ that display nonlinear oscillations of $z_s$ that change sign and spend equal amounts of time with $z_s>0$ and $z_s<0$, from the symmetry broken trajectories, with $E>1$ that oscillate without $z_s$ changing sing.
The eigenstates reflect the structure of the classical trajectories: for $E<1$, the eigenstates are symmetric with $\left<S_s^z\right>=0$, while for $E>1$, nearly degenerate symmetry-broken eigenstates emerge.
%For $E<1$, the eigenstates are symmetric with $\left<S_s^z\right>=0$, and the classical dynamics involve nonlinear oscillations of $z_s$ that change sign and spend equal amounts of time with $z_s>0$ and $z_s<0$.
%While for $E>1$, nearly degenerate symmetry-broken eigenstates emerge, and two corresponding degenerate classical trajectories occur in which $z_s$ oscillates without changing sign.

The singularities of the $E=1$ dynamics are seen in both the classical trajectories and quantum states.
%The dynamics of the classical spin initialized on the separatrix (i.e. $E(\phi_s,z_s)=1$) asymptotically approach the unstable fixed point.
Classically, a spin initialized on the separatrix (i.e. $E(\phi_s,z_s)=1$) shows critically slow dynamics that asymptotically approaches the unstable fixed point.
In the quantum limit, the ESQPT has eigenstates that are localized around the unstable fixed point $\langle\vec{S}_s\rangle= -S\hat{x}$ 
This localization causes singularities in eigenstate observables which in turn causes the sensitivity to the initial state eigenstate distribution $\left|c_n\right|^2$.
To detect memory at the ESQPT, we considered the evolution of different initial states parameterised by $\phi_{s,0}=\phi_s(t=0)$ with $z_s(t=0)$ chosen to satisfy the energy constraint $E(\phi_s,z_s)=1$.
Classically, this means choosing states with ($\phi_s,+\left|z_s\right|$) that lie along the separatrix, while quantum mechanically it means choosing initial states to be the ground state of 
\begin{eqnarray}
    H_0(\phi_s,z_s)=-\vec{h}(\phi_s,z_s)\cdot\vec{S}_s
\end{eqnarray}
such that $\vec{h}$ points in the direction specified by the values of ($\phi_s,+\left|z_s\right|$) that lie  along the separatrix (see Fig.~\ref{fig:classicalOG}).
One then finds that the long-time average quantum observables $\overline{S_s^x}(\phi_{s,0})$ and $\overline{S_s^z}(\phi_{s,0})$ are strongly sensitive to $\phi_{s,0}$, where the bar notation reflects time average and is defined in Eq.~\ref{eq:latetime}.
An example of this is shown in Fig.~\ref{fig:classicalOG}.
There, we show the difference between $\overline{S_s^x}(\pi)$ and $\overline{S_s^x}(0)$ is about $30\%$ for a large spin size of $S=2000$.
It also shows a very slow decay of the sensitivity of $\overline{S_s^x}(\phi_{s,0})$ to $\phi_{s,0}$ with $S$.
This slow decay was shown in Ref.~\cite{Kelly2020} to be logarithmic with $S$ and is because the spin becomes completely classical as $S\rightarrow \infty$.

This type of late-time memory is not expected from the classical integrability present in the model.  The classical model has only two degrees of freedom and thus has only one conserved charge $H_s$.  The conserved charge is the same for the different states discussed above, and so one expects that the different states should have the same late-time observables.
In the classical limit, as $S \rightarrow \infty$, this is true because all the states initialised on the separatrix asymptotically evolve to the unstable fixed point.
In the quantum model, and for any finite $S$, this is not true and we instead find memory of a quantity that is not classically conserved, the phase $\phi_s$.
Note that for initial quantum states parameterised by $\phi_{s,0}$, but with energy constrained to be off the separatrix, $E\neq 1$, the late-time average observables $\overline{S_s^x}(\phi_{s,0})$ and $\overline{S_s^z}(\phi_{s,0})$ do not depend on the initial phase $\phi_{s,0}=\phi_s(t=0)$~\cite{Kelly2020}.

\subsection{Chaotic Dynamics}
\begin{figure}[]
    \centering
    \hspace*{-1.0cm}
    \includegraphics[width=\figsize]{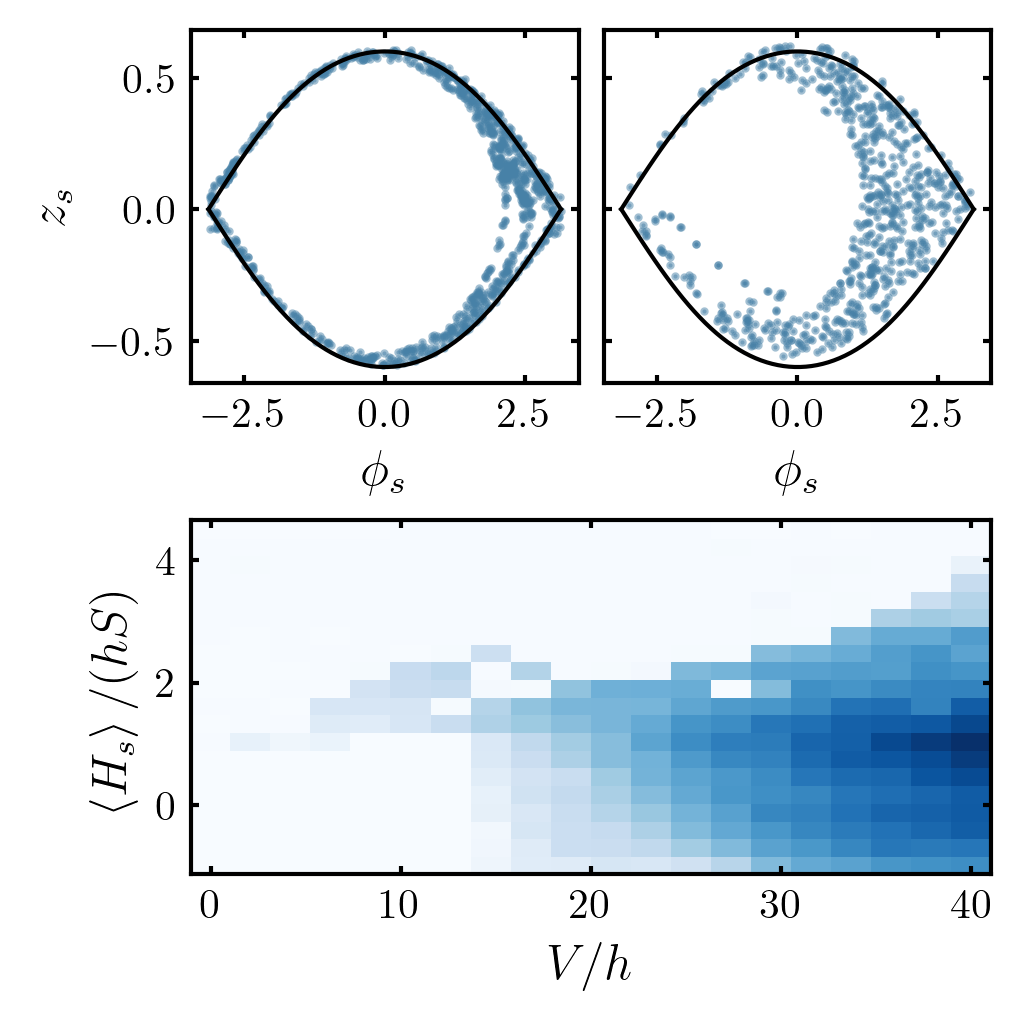}
    \caption{\textit{Top Panels:} These figures show chaotic Poincar\'{e} maps for $V/h=5$ (left) and $V/h=15$ (right). The plane fixing two of the four phase space variables is defined by the total energy constraint and $z_e=-0.1$. The initial state for both left and right figures had $\phi_s(t=0)=0$ and $(\phi_e, z)=(0,0)$, but with different intial $z_s$ fixed by $E(z_s,\phi_s)=1.05$ (left) and $E(z_s,\phi_s)=1$ (right). \textit{Bottom Panel:} This figure shows the Lyapunov exponent versus the initial energy of the system, $E$, and the coupling strength, $V$. 
    For an initial state with $\phi_s=0$ and $z_s$ fixed by the energy constraint $E(\phi_s,z_s)=E$ ($y$-axis) and $(\phi_e,z_e)=(0,0)$.
    The Lyapunov exponent is the exponential rate at which near by trajectories diverge~\cite{DynamicalSystems.jl-2018}, and is positive (darker blue) for chaotic dynamics and zero (white) for regular dynamics.
    The largest Lyapunov exponent found was $\lambda=1.5$ and corresponds to the darkest blue.}
    
    \label{fig:Cchaos}
\end{figure}

In this paper, we investigate how general this type of memory at an ESQPT is.
The previous result occurred in a model in which thermalization was already not expected to occur because of classically regular dynamics and quantum integrability by Bethe ansatz~\cite{MORITA2006337}.
To open the possibility of thermalization we break the integrability present in the model by coupling the system to a non-interacting environment of spins, and consider the effect on the memory phenomenon at $V>0$.

Such a coupling breaks integrability and produces classical chaos.
Similar to the system-only dynamics, a semi-classical description is valid in the large $N$ limit where, in addition to the system spin variables $(\phi_s,z_s)$, the classical dynamics tracks the evolution of the environment collective spin mode $\vec{S}_e$ of size $S=N/2$ via the classical conjugate variables ($\phi_e, z_e$).
As known for a similar model, where the $\vec{S}_e$ is replaced by a harmonic oscillator~\cite{toda2005}, chaos first appears (at small $V/h$) for initial states close to the $E(\phi_s,z_s)=1$ separatrix. 
An example of this chaotic behavior close to the separatrix is demonstrated by the Poincaré maps plotted in Fig.~\ref{fig:Cchaos}. 
These maps indicate chaos by depicting trajectories that cross the Poincaré section at random points near the separatrix.

\begin{figure}[]
    \centering
    \hspace*{-0.5cm}
    \includegraphics[width=\figsize]{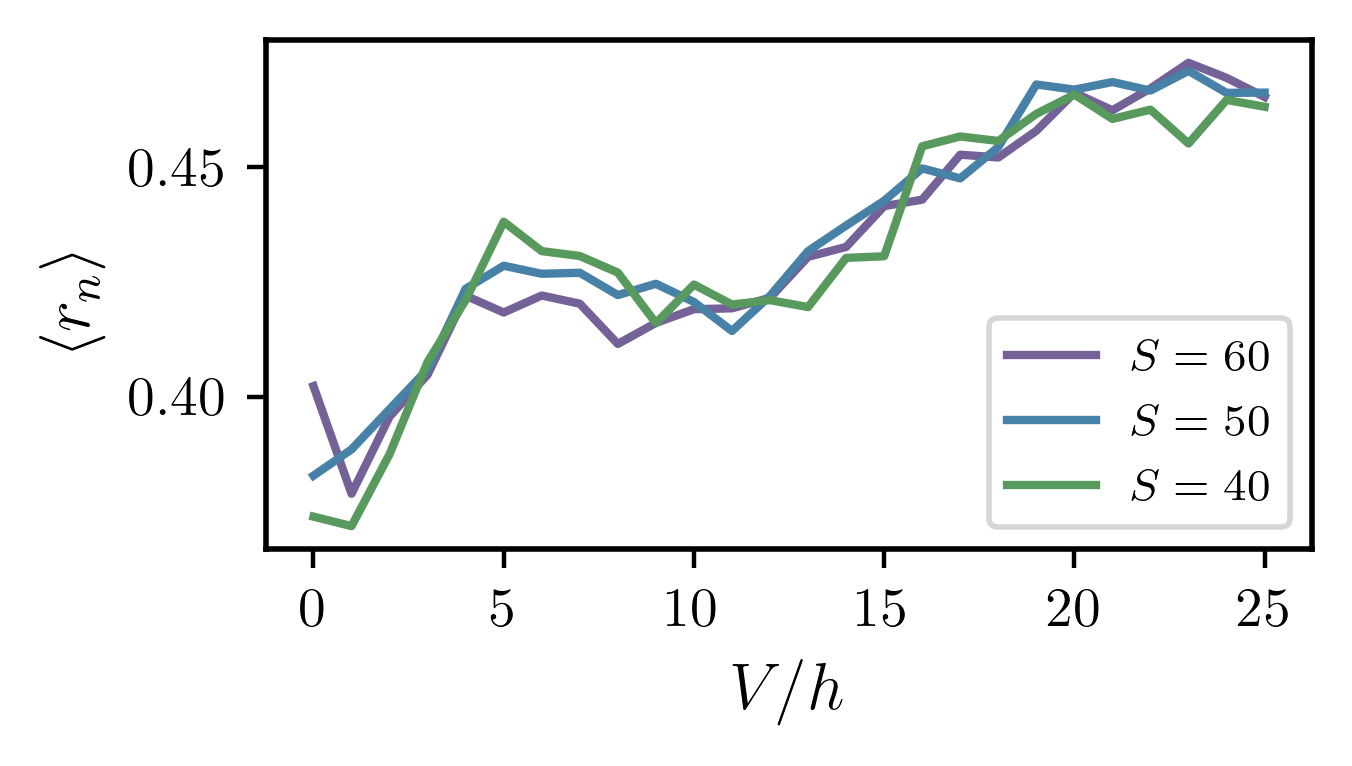}
    \caption{Mean level spacing $\langle r_n\rangle$, for states in the even symmetry sector~($U_{Z2}\left|\psi\right>=\left|\psi\right>$), as a function of system-environment coupling strength $V$ for {$S=40,50$ and $60$}. When $V=0$, the model is integrable and has mean level spacing $\left<r_n\right>\approx0.38$. Upon increasing $V$, the level spacing increases, indicating integrability breaking. While integrability is broken, the level spacing never increases to $\left<r_n\right>=0.54$ which would indicate full quantum chaos.
    }
    \label{fig:mls}
\end{figure}

To understand how chaos emerges as V is increased, we compute~\cite{DifferentialEquations.jl-2017} how quickly trajectories diverge for different initial states.
We consider initial states $\phi_s=0$, $z_e=0$, $\phi_e=0$ and $z_s$ such that $E(\phi_s,z_s)=E$, and compute the Lyapunov exponent $\lambda$, which is shown in Fig~\ref{fig:Cchaos}.
In Fig.~\ref{fig:Cchaos}, we observe that for $V/h<15$, the dynamics is only chaotic near the separatrix, but with varying Lyapunov exponent.
For $V/h>15$ chaos dominates $E<1$, while smaller and smaller regions of $E>1$ remain regular with Lyapunov exponent $\lambda=0$.
This mixture of regular and chaotic trajectories dependent on energy is a typical feature of long range models~\cite{Emary2003} such as the one considered here.

To observe the effect of this chaos in the quantum dynamics, we consider the level statistics.
To do so, we follow~\cite{PhysRevLett.110.084101}, which studied the statistics of $s_n=e_{n+1}-e_n$ where $e_n$ is the $n$-th largest eigenvalue, $H\left|n\right>=e_n\left|n\right>$.
In an integrable system, the distribution of $s_n$ follows a Poisson distribution, while in a chaotic system level repulsion occurs and the distribution follows a Wigner-Dyson distribution.
In Ref.~\cite{PhysRevLett.110.084101}, they argued numerical analysis of $r_n=\min(s_n,s_{n-1})/\max(s_n,s_{n-1})$ is easier and showed that $\left<r_n\right>=0.39$ signals integrability and a Poisson distribution, while $\left<r_n\right>=0.54$ signals level repulsion and Wigner-Dyson statistics.
Since the symmetry breaking leads to a degeneracy, which we do not want to consider in the level statistics, we remove one of the degenerate eigenenergies by only considering eigenstates in one of the symmetry sectors.
Specifically, we consider the $Z_2$ unitary symmetry operator $U_{Z2}$, that takes both $S^z_s\rightarrow -S^z_s$ and $S^z_e\rightarrow -S^z_e$, and only consider states which satisfy both $H\left|n\right>=e_n\left|n\right>$ and $U_{Z2}\left|n\right>=\left|n\right>$.
%Furthermore, since most of the spectrum is classically regular for $V<15$, and we are only interested in the dynamics close to the ESQPT, with system energy $E(\phi_s,z_s)=1$.
%In addition to this filter, we are only interested in the energy levels near the energy of the initial states we are considering.
%Those are states with system energy $E=1$ near the ESQPT, and with environment spins in the ground state of the environment hamiltonian $H_e$: $\langle\vec{S}_e\rangle=S\hat{x}$.
%These states have total energy $\langle H\rangle=0$, so we apply the filter $e_L<e_n<e_u$ where {$e_u=h/S$} and $e_L$ is an arbitrary lower bound to the energy of the eigenstates considered.
%Thus when $V/h$ is small, and $e_L\ll 0$, we only consider the level statistic of the symmetric states with $\langle\vec{S}_x\rangle\sim S\hat{x}$ and system energy $E<1$.
%While when $e_L\rightarrow 0$, we only consider eigenstates relevant for the dynamics of the system and environment initialized near the ESQPT of the $V=0$ model.

\begin{figure}[t]
    \centering
    \hspace*{-0.5cm}
    \includegraphics[width=\figsize]{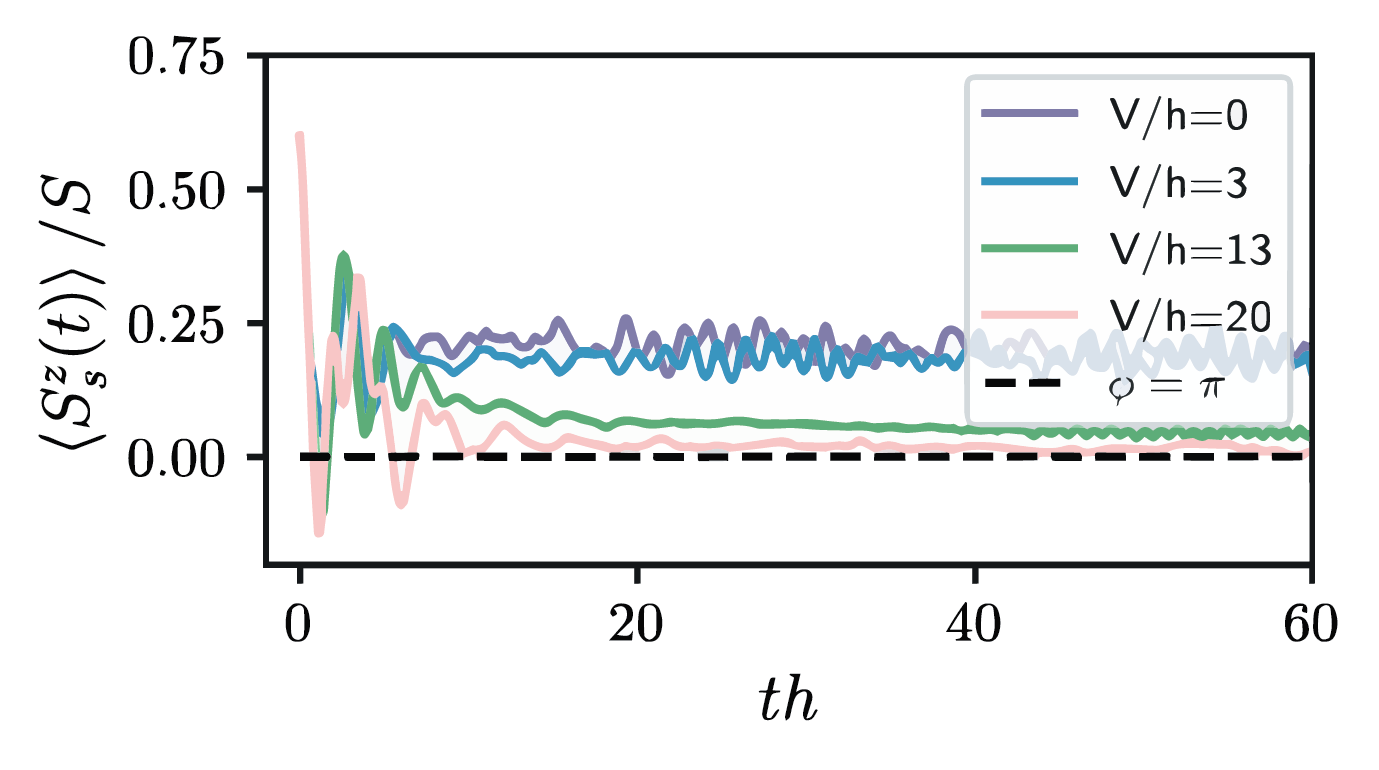}
    \caption{Typical dynamics of $\left<S^z_s(t)\right>$ for the initial state in Eq.~\ref{eq:initstae}, $\phi=0$ (solid lines) and $\phi=\pi$ (dashed-line) and coupling, $V$, shown in the plot. For $\phi=\pi$ the state is $Z_2$ symmetric and $\left<S^z_s\right>=0$ remains symmetric at all times and for all coupling $V$. For small $V$, the steady state of $\left<S^z_s\right>$ distinguishes the $\phi=\pi$ and $\phi=0$ states, while for large values of $V$, this late time memory of $\phi$ is lost in the steady state. }
    \label{fig:dynamics}
\end{figure}

The level statistics for these states are shown in Fig.~\ref{fig:mls}.
There we observe that the amount of level repulsion increases with $V$ but never obtains a completely chaotic value of $\left<r_n\right>=0.54$.
%Furthermore, we find that for $V/h=20$ and $V/h=1$, level repulsion increases closer to the ESQPT transition reflecting the larger Lyapunov exponent near $E\approx1$ as seen in Fig.~\ref{fig:Cchaos}.
%While for $V/h=5$ and $V/h=20$, the level repulsion does not increase as $e_L\rightarrow 0$ reflecting a variability in the amount of chaos as a function of $V/h$ (also observed in the Lyapunov exponent shown in Fig.~\ref{fig:Cchaos}).
Together, the results shown in Fig.~\ref{fig:Cchaos} and Fig.~\ref{fig:mls} demonstrate that, while the full dynamics do not become completely chaotic, the system-environment coupling, $V/h$, breaks the integrability present in the $V=0$ model and creates chaotic classical dynamics (positive Lyapunov exponent) near the ESQPT.

\section{Memory due to an ESQPT}

\begin{figure*}[t!]
    \centering
    \hspace*{-0.5cm}
    \includegraphics[width=\textwidth]{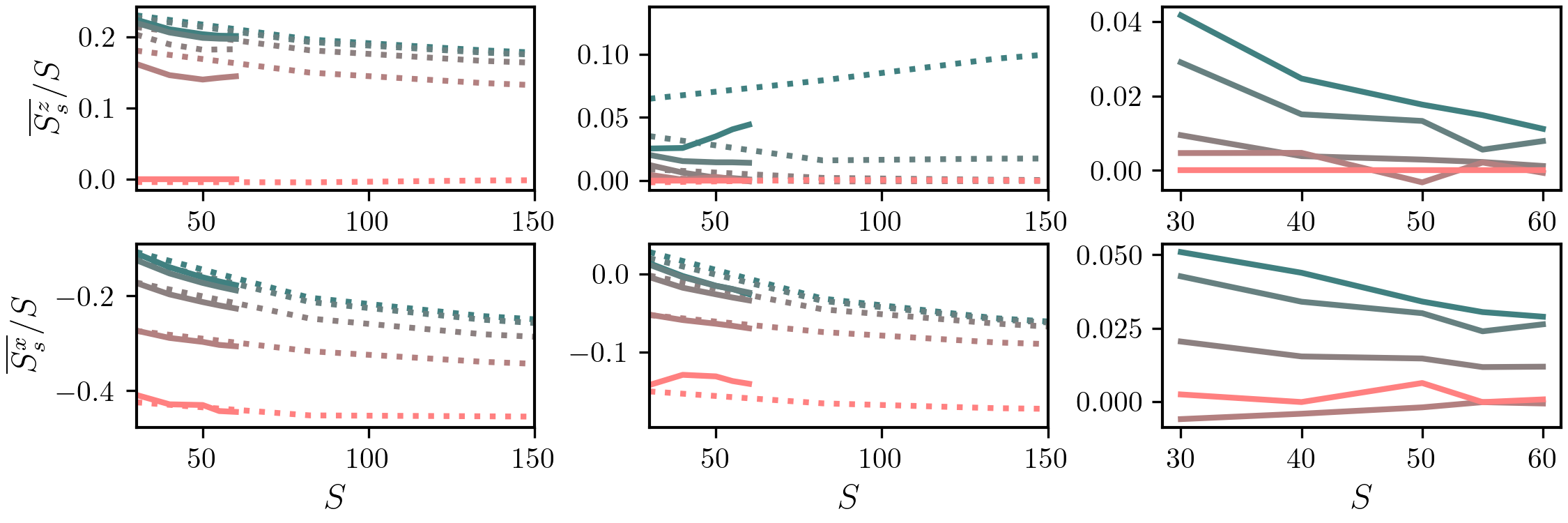}
    \caption{ Dependence of the long time averages $\overline{S_s^x}(\phi_{s,0})$ and $\overline{S_s^z}(\phi_{s,0})$ on S for different initial states $\left|\psi(\phi_{s,0})\right>$ with $\phi_{s,0}$ varying from $\phi_{s,0}=0$ (dark blue) to $\phi_{s,0}=\pi$ (bright pink). The different columns correspond to different coupling strengths $V/h=2$ (left), $V/h=13$ (center), and $V/h=20$ (right). For $V/h=2$, long time memory (sensitivity of the long time observable to the initial system phase $\phi_{s,0}$) is observed even at large values of $S$ similar to the $V/h=0$ case shown in Fig.~\ref{fig:classicalOG}.  While for $V/h=20$ sensitivity of initial conditions is quickly loss as the spin size $S$ increase.  
        Exact calculations are shown by solid lines, while semi-classical observables calculated via TWA (with sampling error $\Delta O/S\approx 0.025$)  are shown by dashed lines.
        We leave out TWA calculations for $V/h=20$ because observable values are smaller than the sampling error.
    }
    \label{fig:OvsJ}
\end{figure*}
To investigate the robustness of the ESQPT memory effect to the integrability breaking coupling with the environment, we consider the dynamics of an initial product state between system and environment:
\begin{eqnarray}\label{eq:initstae}
    \left|\psi(t=0)\right>=\left|\psi(\phi_{s,0})\right>=\left|\phi_{s,0},z_{s,0}\right>\otimes\left|\psi_e\right>
\end{eqnarray}
where the initial states of the environment and system, $\left|\psi_e\right>$ and $\left|\phi_{s,0},z_{s,0}\right>$, are spin coherent states polarized in the $\hat{x}$ direction and $(\phi_{s,0},z_{s,0})$ direction respectively:
\begin{eqnarray}
    S_e^x\left|\psi_e\right>&=&-S\left|\psi_e\right> \\ \nonumber
    H_0(\phi_{s,0},z_{s,0})\left|\phi_{s,0},z_{s,0}\right>&=&-S\left|\phi_{s,0},z_{s,0}\right>.
\end{eqnarray}
Importantly, the $\hat{z}$-polarization of the system spin, $z_{s,0}$, is chosen dependent on $\phi_{s,0}$ such $E(\phi_s,z_s)=1$ and the system starts at the ESQPT of the $V=0$ model.
Notice that the different initial states considered, $\left|\psi(\phi_{s,0})\right>$, are parameterized only by the angle of the system spin from the $\hat{x}$-axis, and that they all have the same total energy $\left<H\right>=0$, and system energy $E(\phi_s,z_s)=1$, that is independent of the choice of $\phi_{s,0}$.
This is because 1) the system-environment coupling energy is zero, $\left<H_v\right>=0$, for all initial states; 2) the environment energy $\left<H_e\right>$ is trivially independent of system state parameter $\phi_{s,0}$, and 3) the system energy $\left<H_s\right>=E(\phi_s,z_s)Sh_s$ is fixed by the constraint $E(\phi_s,z_s)=1$.
Therefore if the long time observables $\overline{S_s^x}(\phi_{s,0})$ and $\overline{S_s^z}(\phi_{s,0})$ depend on $\phi_s$, then they will indicate initial state memory and sensitivity to the initial state amplitudes, $c_n$, beyond what can be expected from total energy conservation.

The typical dynamics following the quench are shown in Fig.~\ref{fig:dynamics}, and show how, for small $V$, the steady state dynamics distinguish between the initial state at $\phi=0$ and $\phi=\pi$.
    Since, at finite size, the late time dynamics generically show temporal fluctuations~\cite{khripkov_temporal_2013,Alessio2016}, we study late time memory of the initial state by considering the time average expectation values shown in Fig.~\ref{fig:OvsJ}.
There we see that for $V/h=2$ states close to the unstable fixed point, $\phi_s=\pi$, have late-time expectation values,  $\overline{S_s^x}(\phi_{s,0})$ and $\overline{S_s^z}(\phi_{s,0})$, close to the value of the unstable fixed point (i.e. $\left<S_s^x\right>=-1$ and $\left<S_s^z\right>=0$), while for states far from the unstable fixed point, we see see late-time expectations values depart from the $\left<S_s^x\right>=-1$ and $\left<S_s^z\right>=0$, with a larger difference the further $\phi$ is from $\pi$.
This confirms that ESQPT memory effect is robust for at least one value of $V$.
While for $V/h=20$, we see a loss of memory, with late-time observables approaching the same value for different initial phases of the system spin, $\phi_s(t=0)=\phi_{s,0}$.

\begin{figure}[]
    \centering
    \hspace*{-0.5cm}
    \includegraphics[width=\figsize]{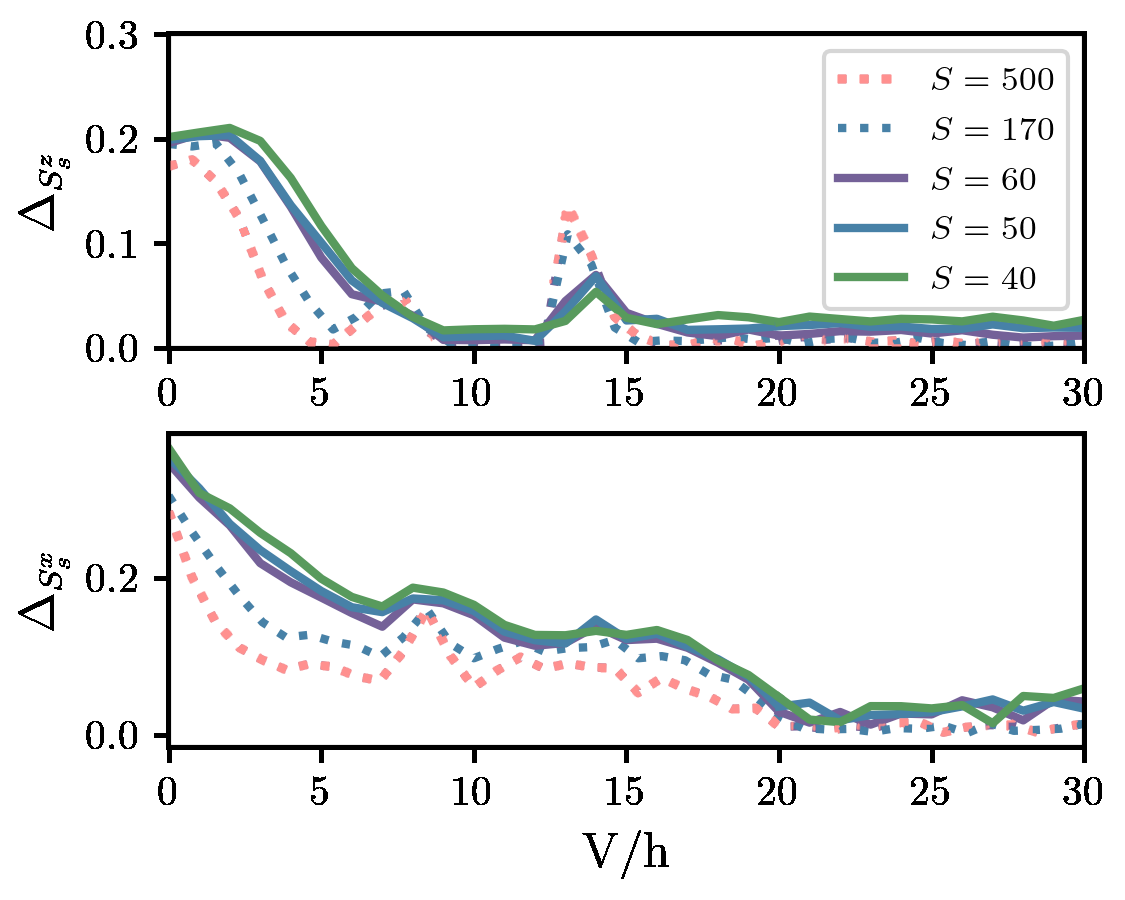}
    \caption{
            Memory quantifiers $\Delta_{S^z_s}$ and $\Delta_{S^x_s}$ for the long time averages $\overline{S_s^x}(\phi_{s,0})$ and $\overline{S_s^z}(\phi_{s,0})$ as a function of the system-environment
         coupling $V$. 
            Exact calculations are shown by solid lines, while TWA calculations  are shown by dashed lines.
            The bottom figure indicates memory of $\phi_{s,0}$ is maintained until $V/h=20$ in the long time average of $S^x_s$, while the top figure indicates it is lost earlier  for $V/h\sim 16$ in the long time average of $S^z_s$.
            The weakening of memory with increasing spin size, $S$ is due to the classicality of the collective spin in the $S\rightarrow \infty$ limit, and is consistent with the logarithmic scaling of $S$ found in Ref.~\cite{Kelly2020}.
    }
    \label{fig:DvsV}
\end{figure}

These plots show detailed structure of the late-time memory as a function of spin size $S$ for different values of $\phi_{s,0}$, but to observe the explicit dependence of the late-time memory on the system environment coupling $V$, we need to compress the information shown in these plots.
This can be done by instead considering the following memory quantifier:
\begin{eqnarray}
    \Delta_O = \left(\max_{\phi_{s,0}}\overline{O}(\phi_{s,0})-\min_{\phi_{s,0}}\overline{O}(\phi_{s,0})\right)/S,
\end{eqnarray}
where the maximum and minimum are over different initial phase angles, $\phi_{s,0}$, for the initial states defined in Eq.~\ref{eq:initstae}.
When there is no late time memory, the late time dynamics are insensitive to the initial state phase and $\Delta_O=0$.
On the other hand, if $\Delta_O$ is large, it implies that two different initial states can be easily distinguished by the late time average observables $O$, and that the $O$ remember the phase, $\phi_{s,0}$, of the initial state. 
This memory quantifier is shown in Fig.~\ref{fig:DvsV}, and demonstrates that quantum memory due to the ESQPT is present for $V/h<20$, but is lost for $V/h>20$.
This demonstrates that this type of memory is robust to integrability breaking perturbations up to large perturbation strength, even for $V/h>15$ when the model demonstrates chaotic dynamics~(see Fig.~\ref{fig:Cchaos} which shows positive Lyapunov exponent).
This is in contrast with the memory occurring in perturbed integrable models, where memory of initial conditions is conserved in quasi-conserved charges only up to some prethermal plateau, after which the dynamics relaxes to a thermal state with no memory at much longer times~\cite{Mori_2018,Matteo2013}.  Here we see persistent late time memory with no indication of a prethermal plateau.

\begin{figure}[]
    \centering
    \hspace*{-0.5cm}
    \includegraphics[width=\figsize]{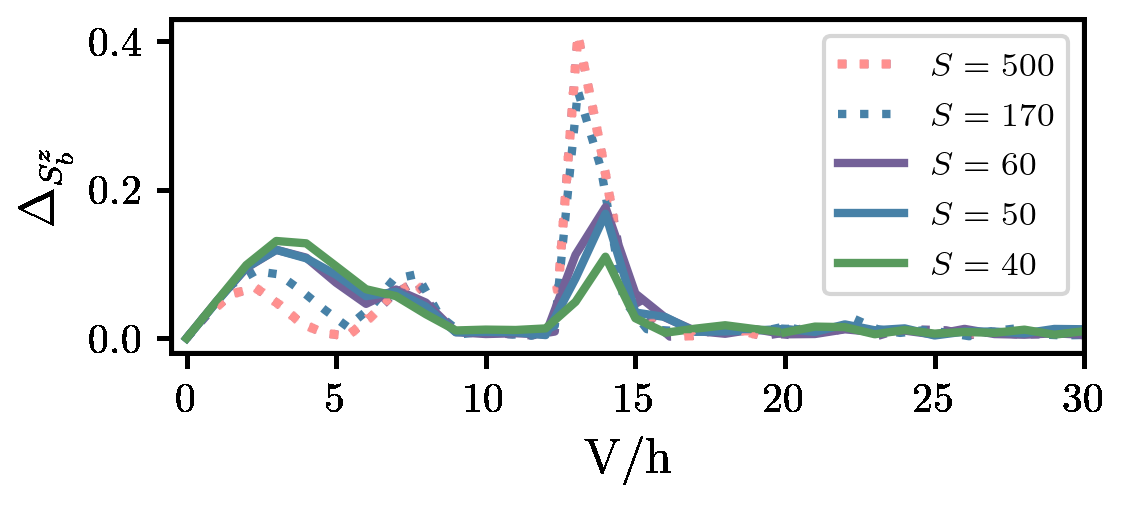}
    \caption{
         This figure shows the memory quantifier $\Delta_{S^z_e}$ for the long time average of environment $z$-polarization $S^z_e$ as a function of the system-environment coupling $V$.
        It indicates that for $0<V/h<16$ a type of memory proximity effect occurs, where the initial state of the system spins is remembered by the long time average of the environment spins.
        Exact calculations are shown by solid lines, while TWA calculations  are shown by dashed lines.
        %\textit{Bottom Panel}: Long time average of the environment energy $\left<H_e\right>/(hS)=S^x/S$ as a function of the system-environment coupling $V$.  It shows that as the coupling is increased, the environment is more strongly excited.
    }
    \label{fig:S2vV}
\end{figure}

We also find that the ESQPT memory has a type of proximity effect on the environment, where the long time observables for the environment become sensitive to the initial state of the system.
This is demonstrated in $\Delta_{S^z}$ shown in Fig.~\ref{fig:S2vV}.
There, at $V=0$, the state of the environment is insensitive to $\phi_s(t=0)$, while as $V$ increases the environment becomes more sensitive, until $V/h=4$.  
This can be understood by considering how the environment is excited by the system differently between the $\phi_{s,0}=0$ and $\phi_{s,0}=\pi$ states.
For initial states with $\phi_{s,0}=0$, the $z$-component of the system spin is large and remains positive at long times.
The environment spins therefore see an effective field, $+\frac{V}{N}\left<S^{z}_s\right>S^{z}_e$, which directly drives environment spins rotating around a field pointing in the $-h\hat{x}+\frac{V}{N}\left<S^{z}_s\right>\hat{z}$ direction.
Thus for initial states with $\phi_{s,0}$ the environment spins rotate and obtain a positive $\hat{z}$-polarization $\left<S^{z}_e\right>>0$.
While for $\phi_{s,0}=\pi$, $\left<S^z_s\right>$ remains $0$, the effective field remains $0$, and only symmetric quantum fluctuations of $S^{z}_s$ can excite the environment via the system environment coupling.
Thus, the environment is excited symmetrically and $\left<S^{z}_e\right>=0$ at late times.

Finally, we would like to point out an apparent memory peak occurring at $V/h\approx13$.
At this peak, the late time quantum memory increases with system size in both the system and environment observables, $\Delta_{S^x_s}$ and $\Delta_{S^x_e}$, as shown in Figs.~\ref{fig:DvsV}
and ~\ref{fig:S2vV}.
This is distinct from the $V=0$ case, in which late time memory decreases in the large spin limit as the spin becomes more classical.
A detailed analysis of this feature is left for future work, but it appears to be related to when the ground state breaks the Z$_2$ symmetry at a similar coupling strength (see discussion below and Fig.~\ref{fig:degenlevels}).

\section{Loss of the ESQPT Memory Effect at Large $V$}

\begin{figure}[]
    \centering
    \hspace*{-0.5cm}
    \includegraphics[width=\figsize]{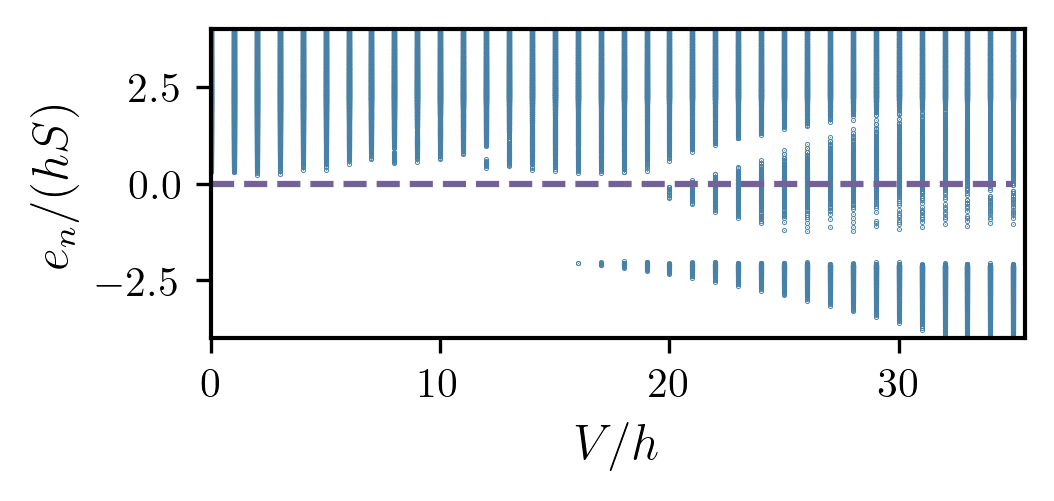}
    \caption{This figure is a scatter plot of the degenerate energies for different values of $V/h$. We consider two energy levels degenerate if $\left|e_{n}-e_{n+1}\right|<10^{-7}$, where $e_n$ is the n$^{th}$ largest eigen energy: $H\left|e_n\right>=e_n\left|e_n\right>$.  It shows that for $V/h<15$, degeneracy only occurs above $e_n=0$, while for $V/h>20$ the ESQPT responsible for memory at $V=0$ shifts to lower energy. The purple dashed line marks the $V=0$ ESQPT at $e_n/(hS)=0$.}
    \label{fig:degenlevels}
\end{figure}

Above, we have seen that the memory at the ESQPT of the $V=0$ model is robust to chaos, but for $V/h>20$ it appears to be lost.
This seems surprising because there is still an unstable fixed point at $(\phi_s,z_s)=(\pi,0)$ which could create localized eigenstates in the spectrum as was the case in Ref.~\cite{Kelly2020}.
This is because the unstable fixed point is a singular point in a larger classical phase space that is not guaranteed to show up in the statistical theory or eigenstate observables.
This is in contrast to a phase transition, which is a singularity in the statistical theory and more strongly guarantees singularities in $O(e_n)$.
Therefore we should check if there is still an ESQPT at an energy $\left<H\right>=0$ to observe if the ESQPT memory effect should remain at strong coupling $V$.

To do so, we use the fact that, in the large $S$ limit, there are degenerate eigenstates in the symmetry broken phase of a Z$_2$ symmetry.
This is because in the symmetry broken phase the Z$_2$ symmetry operator, $U_{Z2}$,  acts non-trivially on the symmetry broken eigenstate $U_{Z2}\left|e_n,0\right>= \left|e_n,1\right>\neq\left|e_n,0\right>$.
Furthermore, since the Hamiltonian is symmetric and $[H,U_{Z2}]=0$, the eigenstate $\left|e_n,1\right>$ is degenerate with the eigenstate $\left|e_n,0\right>$.
Therefore $0$ and $1$ label the two distinct degenerate eigenstates.

In Fig.~\ref{fig:degenlevels}, we have plotted all the energies of the degenerate energy levels.
There, we see that for $V/h<15$, level degeneracy, and thus symmetry breaking, occur for $\left<H\right> > 0$, indicating that the zero-coupling ESQPT occurring $\left<H\right>=0$ is maintained for significantly large $V/h$.
Then at intermediate $15<V/h<20$, we find an additional transition occurring at $\left<H\right>=-Sh$, but with the transition at $\left<H\right>=0$ remaining.
Below this addition transition, the lowest energy states in the spectrum break the Z$_2$ symmetry.
While for $V/h>20$, the $\left<H\right>=0$ transition decreases in energy.
Thus, there is no longer an ESQPT at the energy of the initial states we considered above, and thus no longer any obstruction to memory loss due to singularities in eigenstate observables.

It is possible that the memory effect remains for large system environment coupling, $V$, but now at the two eigenstate phase transitions for $\left<H\right> < 0$.
Investigation into this possibility is currently beyond the scope of this work because it requires a better understanding of these phase transitions.
In particular, the energy at which the new ESQPTs occurs needs to be precisely identified along with initial states that can be used to test the memory effect.

%Unfortunately, our model does not host an ESQPT as a function of energy for large coupling.  

\section{Discussion}

We have studied the effect of ESQPT memory in a weakly chaotic Z$_2$ symmetric model.
We show that the late time memory effect found in the integrable model of Ref.~\cite{Kelly2020}~(summarized in Fig.~\ref{fig:classicalOG}) is stable to an integrability breaking system-environment coupling and in the presence of a chaotic classical limit.
We find that initial state properties, which are not conserved by the dynamics, are remembered at late times when the initial state energy is at the ESQPT breaking the Z$_2$ symmetry.
Furthermore, we see that at small, but non-perturbative, system-environment coupling, a memory proximity effect is induced, in which the memory of initial conditions is also seen in the non-interacting environment.
At large system-environment coupling, the ESQPT moves to a lower energy $\left<H\right><0$, and the memory due to the ESQPT is no longer observed by initial states at $\left<H\right>=0$.

This work more firmly establishes the effect of ESQPT memory in a non-integrable model.
Future work could study this effect in a finite dimensional model, where chaos, and a clear thermal phase transition are more cleanly available for study. 
The main difficulty in this direction is finding a model that is tractable for quantum dynamics.
Models in 1D sometimes offer this, but limit the ability of symmetry breaking, while long-range (infinite dimension) models suppress local fluctuations and generate regular structure in collective modes~\cite{Kelly2021,Lerose2019,Kac1963,CAMPA200957,Kastner2013,Kastner2011,Worm_2013}.
Another possibility is to study other symmetry groups beyond Z$_2$, such as in various $O(N)$ fully connected models that are known to host an ESQPT~\cite{Cejnar_2021}.

Another interesting possibility is suggested by works studying dynamical phase transitions~\cite{Smale2019,zhang2017,Hwang2015,Alessandro2018,Lerose2018,Lerose2019a, Jeremy2020}.
In particular, Ref.~\cite{Titum2020} showed that critical exponents in a quench to a model with a ground state phase transition can be sensitive to the initial state. 
The results observed there do not demonstrate long-time memory because the initial states had different energy, but it is possible that the initial-state memory here also shows up in critical exponents.
More generally, the contradiction between thermalization and phase transitions discussed in the introduction has a simple but fundamental
implication on the universality classes of non-equilibrium phase transitions in closed quantum systems: if thermalization is not possible for the reasons discussed above, distinct universality classes can be expected for the phase transition following quenched dynamics and the one occurring at thermal equilibrium.

%\begin{acknowledgments}
{\bf Acknowledgments.} During preparation of this manuscript, we became aware of reference~\cite{sinha2022}, which studies the quantum scars that form near the ESQPT in a similar model. We are grateful to Riccardo J. Valencia-Tortora and Jamir Marino for their useful comments on the manuscript. This work was supported in part by the National Science Foundation (NSF) RAISE-TAQS under Award Number 1839153 (SWT). Computations were performed using the computer clusters and data storage resources of the HPCC at UCR, which were funded by grants from NSF (MRI-1429826) and NIH (1S10OD016290-01A1).
This work (SPK) has been funded by the Deutsche Forschungsgemeinschaft (DFG, German Research Foundation) - TRR 288 - 422213477 (project B09), TRR 306 QuCoLiMa (”Quantum Cooperativity of Light and Matter”), Project-ID 429529648 (project D04) and in part by the National Science Foundation under Grant No. NSF PHY-1748958 (KITP program ‘Non-Equilibrium Universality: From Classical to Quantum and Back’). 

%\end{acknowledgments}

\bibliography{refs.bib}
 
\end{document}